\documentstyle[12 pt]{article}
\begin {document}
\title{
  S-matrix for a  spin $1\over 2$ particle in a Coulomb +Scalar potential }
\author{Arvind Narayan Vaidya$^{\star}$\\
Luiz Eduardo Silva Souza\\
Instituto de F\'\i sica - Universidade Federal do Rio de Janeiro \\
Caixa Postal 68528 - CEP 21945-970, Rio de
Janeiro, Brazil
\\
} 
\date{}
\maketitle
\begin{abstract}
The S-matrix for a spin 1/2 particle in the presence of a potential which is the sum of the Coulomb potential $ V_c=-A_1/r$ and a Lorentz scalar potential  $V_s=-A_2/r$ is calculated. 
\end{abstract}
\vskip 2.5 cm
\noindent
PACS numbers:03.65.Pm,11.80.-m
\vskip 1.5 cm 
\noindent
$^{\star}$ {e-mail:vaidya@if.ufrj.br}\nonumber\\
\pagebreak
\vskip 1.0 cm
\noindent
{\bf 1.Introduction}
\par
 We consider a spin ${1\over 2}$ particle in a potential which is the sum of the Coulomb potential $ V_c=-A_1/r$ and a Lorentz scalar potential  $V_s=-A_2/r$.The scalar potential is added to the mass term in the Dirac equation and may be interpreted as an effective position dependent mass.If the scalar potential is assumed to be created by the exchange of massless scalar mesons, it  has the form $V_S= - {A_2\over r}$.
\par
Exact solutions for the bound states in this mixed potential can be obtained by separation of variables [1,2]. In this paper we consider the scattering problem for such a potential which does not seem to have been treated in the literature. We calculate the phase shifts by the conventional technique and show how the scattering problem can also be solved algebraically. 
\par
This paper is organized as follows.In section 2 we separate variables in the Dirac equation obtaining the radial equations,in section 3 we solve the radial equations for the scattering problem and calculate the phase shifts,in section 4 we apply an algebraic technique to obtain the phase shifts.Section 5 contains the conclusions.
\vskip 1.0 cm
\noindent 
{\bf 2. Separation of variables in the Dirac equation}
\par
The time independent Dirac equation in the presence of the mixed potential may be written as 
\begin{eqnarray}
( {{\mbox{\boldmath $\alpha$}}\cdot{\bf p}} +\beta (M-{A_2\over r})-(E+ {{A_1}\over r}))\Psi(\bf x)=0
\end{eqnarray}
where ${\bf p}=-i {\partial\over {\partial \bf x}}$ and $r=|\bf x|$.
To separate variables we write $\Psi(\bf x)$ in terms of two component spinors 
\begin{equation}
\Psi = \pmatrix{\phi\cr\chi\cr}.
\end{equation}
The two component angular solutions are eigenfunctions of $J^2,J_z,L^2,S^2$ and are of two types 
\begin{equation}
{\phi^{(+)}}_{j,m} = \pmatrix {{({{l+{1\over 2}+m}\over {2l+1}})^{1\over 2}}Y_{l,{m-{1\over 2}}}\cr {({{l+{1\over 2}-m}\over {2l+1}})^{1\over 2}}Y_{l,{m+{1\over 2}}}},
\end{equation}
for $ j = l+{1\over 2}$ and
\begin{equation}
{\phi^{(-)}}_{j,m} = \pmatrix {{({{l+{1\over 2}-m}\over {2l+1}})^{1\over 2}}Y_{l,{m-{1\over 2}}}\cr -{({{l+{1\over 2}+m}\over {2l+1}})^{1\over 2}}Y_{l,{m+{1\over 2}}}},
\end{equation}
for $ j = l-{1\over 2}$.
\par
In the above basis one can verify that
\begin{equation}
J^2{\phi^{(\pm)}}_{j,m}=j(j+1){\phi^{(\pm)}}_{j,m}.
\end{equation}
\begin{equation}
(1+{{\mbox{\boldmath $ \sigma$}}\cdot{\bf L}}){{\phi^{(\pm)}}_{j,m}}= -\kappa{{\phi^{(\pm)}}_{j,m}},
\end{equation}
where $\kappa = \pm (j+{1\over 2})$ for $j=l\mp{1\over 2}$.
Next,we put
\begin{equation}
\Psi=\pmatrix{{iG_{lj}\over r}{\phi^l}_{jm}\cr {F_{lj}\over r}{{\mbox{\boldmath $\sigma$}}\cdot{\bf n}}{\phi^l}_{jm}\cr},
\end{equation}
where ${\bf n}={{\bf r}\over r}$ and
\begin{equation}
{\phi^l}_{jm}={{\phi^{(\pm)}}_{j,m}},
\end{equation}
for $j=l\pm{1\over 2}$.
Defining the Dirac operator
\begin{equation}
K={\gamma}^0(1+{{\mbox{\boldmath $\Sigma$}}\cdot{\bf L}})
\end{equation}
we have
\begin{equation}
K\Psi=-\kappa\Psi
\end{equation}
where we have used the relation 
\begin{equation}
\lbrack1+{{\mbox{\boldmath $\sigma$}}\cdot{\bf L}},{{\mbox{\boldmath $\sigma$}}\cdot{\bf n}}\rbrack_{+}=0.
\end{equation}
Next,using the relations
\begin{equation}
{{\mbox{\boldmath $\sigma$}}\cdot{\bf p}}{f(r)\over r}{\phi^l}_{jm}=-{i\over r}({df\over dr}+{\kappa f\over r}){{\mbox{\boldmath $\sigma$}}\cdot{\bf n}}{\phi^l}_{jm},
\end{equation}
and
\begin{equation}
{{\mbox{\boldmath $\sigma$}}\cdot{\bf p}}{{\mbox{\boldmath $\sigma$}}\cdot{\bf n}}{f(r)\over r}{\phi^l}_{jm}=-{i\over r}({df\over dr}-{\kappa f\over r}){\phi^l}_{jm},
\end{equation}
we get the radial equations
\begin{eqnarray}
{dG_{lj}\over dr}+{\kappa\over r}G_{lj}-(E+M-{A_2\over r}+{A_1\over r})F_{lj}&=&0,\nonumber\\
{dF_{lj}\over dr}-{\kappa\over r}F_{lj}
+(E-M+{A_2\over r}+{A_1\over r})G_{lj}&=&0.
\end{eqnarray}
Next, let $\rho=kr$ where $k^2=E^2-M^2$
Then the radial equations take the form (we omit the indices $l,j$)
\begin{eqnarray}
{dG\over d\rho}&=&({{E+M}\over k}+{{A_1-A_2}\over \rho})F-{\kappa\over \rho}G,\nonumber\\
{dF\over d\rho}&=& -({{E-M}\over k}+{{A_1+A_2}\over \rho})G+{\kappa\over \rho}F.
\end{eqnarray}
{\bf 3. Direct calculation of the phase shifts}
\par
In this section we solve the radial equation by using a technique similar to that used by Lin [3].
Defining the functions $u(\rho)$ and $v(\rho)$ by
\begin{eqnarray}
G&=& {1\over 2}e^{i\rho}(u+v),\nonumber\\
F&=&{i\over 2}({{E-M}\over {E+M}})^{1\over 2}e^{i\rho}(u-v).
\end{eqnarray}
we get
\begin{eqnarray}
{du\over d\rho}&=&{iE\gamma\over k\rho}u+{1\over \rho}(-\kappa+{iM{\gamma}'\over k})v,\nonumber\\
{dv\over d\rho}&=&-({\kappa\over \rho}+i{M{\gamma}'\over k\rho})u-i{\gamma E\over k\rho}v -2iv,
\end{eqnarray}
where $\gamma=A_1+{M\over E}A_2$ and ${\gamma}'=A_1+{E\over M}A_2$.
Eliminating $v$ we get
\begin{equation}
\rho{{d^2u}\over d\rho^2}+(1+2i\rho){du\over d\rho} +({2E \gamma\over k}-{\lambda^2\over \rho})u=0,
\end{equation}
where $\lambda={\kappa\over |\kappa|} (\kappa^2-{A_1}^2+{A_2}^2)^{1\over 2}$.
Let
\begin{equation}
u(\rho)={\rho}^{|\lambda|}w(\rho).
\end{equation}
Then  $w(\rho)$ satisfies the equation
\begin{equation}
\rho{{d^2}w\over d\rho^2}+(2|\lambda|+1+2i\rho){dw\over d\rho}+2i(|\lambda|-{iE\gamma\over k})w=0.
\end{equation}
Restoring the indices $l,j$ we get
\begin{equation}
u_{lj}=a_{lj}{\rho^|\lambda|}\Phi(|\lambda|-{iE\gamma\over k},2|\lambda|+1,-2i\rho),
\end{equation}
where $a_{lj}$ is a constant and $\Phi(a,b,z)$ is the confluent hypergeometric function.
Using equation (17)it is easy to show that 
\begin{equation}
v_{lj}=a_{lj}{{|\lambda|-{iE\gamma\over k}}\over {-\kappa+{iM{\gamma}'\over k}}}{\rho}^{|\lambda|}\Phi(|\lambda|-{iE\gamma\over k}+1,2|\lambda|+1,-2i\rho).
\end{equation}
For $r\to\infty$ we get 
\begin{eqnarray}
G^{out}&=&{a_{lj}\over 2^{|\lambda|+1}}e^{-{E\gamma\pi\over 2k}}e^{ikr+i{E\gamma\over k}\ln2kr}{\Gamma(2|\lambda|+1)\over \Gamma(1+|\lambda|+i{E\gamma\over k})},
\nonumber\\
G^{in}&=&{a_{lj}\over 2^{|\lambda|+1}}e^{-{E\gamma\pi\over 2k}}e^{-ikr-i{E\gamma\over k}\ln2kr}{\Gamma(2|\lambda|+1)\over \Gamma(1+|\lambda|-i{E\gamma\over k})}\times\nonumber\\&e^{i\pi|\lambda|}&{{|\lambda|-i{E\gamma\over k}}\over   {-\kappa+i{M{\gamma}'\over k}}}.
\end{eqnarray}
Thus the phase shifts are given by 
\begin{equation}
e^{2i\delta_{l,j}(k)}= {{-\kappa+i{M{\gamma}'\over k}}\over {|\lambda|-i{E\gamma\over k}}}{\Gamma(1+|\lambda|-i{E\gamma\over k})\over 
\Gamma(1+|\lambda|+i{E\gamma\over k})}e^{-i\pi\lambda}.
\end{equation}
\vskip 1.0 cm
\noindent

{\bf 4. Algebraic calculation of the phase shifts}
\vskip 1.0 cm
\noindent
In this section we apply an algebraic technique to calculate the phase shifts.This technique was used earlier by  Alhassid, G\"ursey and Iachello [4] for the relativistic Coulomb problem.
\par
With $\Phi=\pmatrix{G_{lj}\cr F_{lj}\cr}$ we get
\begin{eqnarray}
\lbrack {d\over dr} +{1\over r}(\kappa \rho_3+A_2\rho_1-iA_1\rho_2)-M\rho_1-iE\rho_2\rbrack \Phi = 0,
\end{eqnarray}
where ${ \rho_i} $ are the Pauli matrices.
\par 
  The potential matrix $\Lambda=A_2\rho_1-iA_1\rho_2+\kappa \rho_3$ may be diagonalized by using the result
\begin{equation}
e^{i\beta\rho_2}e^{-\alpha\rho_1}\Lambda e^{\alpha\rho_1}e^{-i\beta\rho_2} =\lambda\rho_3,
\end{equation}
where
\begin{eqnarray}
\tanh 2\alpha&=&{A_1\over \kappa},\nonumber\\
\tan 2\beta &=& {A_2\over \kappa'},\nonumber\\
\kappa'&= &\epsilon(\kappa)(\kappa^2-{A_1}^2)^{1\over 2},\nonumber\\
\lambda &= &\epsilon(\kappa){({\kappa'}^2+{A_2}^2)^{1\over 2}},
\end{eqnarray}
where $\epsilon(\kappa)={\kappa\over |\kappa|}$.
Defining
\begin{equation}
\Phi = e^{\alpha\rho_1}e^{-i\beta\rho_2} {\hat\Phi},
\end{equation}
we get
\begin{eqnarray}
\lbrack{d\over dr}+ {\lambda\over r}\rho_3+i{\bf k}\cdot{\mbox{\boldmath $\rho$}}\rbrack{\hat\Phi}=0,
\end{eqnarray}
where
\begin{eqnarray}
ik_1&=&{E{A_1}{A_2}-M{\kappa'}^2\over \kappa'\lambda},\nonumber\\
ik_2&=&-iE{\kappa\over \kappa'},\nonumber\\
ik_3&=&-{(MA_2+EA_1)\over \lambda}.
\end{eqnarray}
One can verify that
\begin{equation}
E^2={\bf k}^2 + M^2.
\end{equation}
\par
Multiplying on the left by the operator ${d\over dr} -{\lambda\over r} \rho_3 -i{\bf k}\cdot{\mbox{\boldmath $\rho$}}$ we get the second order equation 
\begin{equation}
\lbrack{d^2\over dr^2} -{\lambda\over r^2}\rho_3-{\lambda^2\over r^2}+{2E\gamma\over r}+ k^2\rbrack \hat\Phi=0,
\end{equation}
where
\begin{equation}
\gamma=A_1+{M\over E}A_2.
\end{equation}
For $r\to\infty$ omitting the $1\over r$ terms in equation (25) we have two types of free particle solutions which
satisfy the conditions 
\begin{eqnarray}
(M\rho_1+iE\rho_2)\Phi^{in}&=&-ik\Phi^{in},\nonumber\\
(M\rho_1+iE\rho_2)\Phi^{out}&=&ik\Phi^{out},
\end{eqnarray}
where $k^2=E^2-M^2$.The explicit form of these solutions is 
\begin{eqnarray}
\Phi^{in}&= &g^{in}\pmatrix{1\cr{ik\over E+M}\cr}e^{-ikr},\nonumber\\
\Phi^{out}&=& g^{out}\pmatrix{1\cr{-ik\over E+M}\cr}e^{ikr}.
\end{eqnarray}
To calculate the scattering matrix we need to find the ratio $g^{out}\over g^{in}$.
\par 
Next,equations (29)and (30) give,in the limit $r\to\infty$ 
\begin{equation}
{g^{out}\over g^{in}}={{\hat g}^{out}\over {\hat g}^{in}}{{ik_{-} S_{11}-ik_ 3S_{12}-ik S_{12}}\over {ik_{-} S_{11}-ik_ 3S_{12}+ik S_{12}}},
\end{equation}
where $S=e^{\alpha\rho_1}e^{-i\beta\rho_2}$.
After considerable effort we get
\begin{eqnarray}
 (ik_{-} S_{11}-ik_ 3S_{12})&=&-{{E+M}\over {A_1-A_2}}(\kappa+\lambda){{(A_1({\kappa}'+\lambda)-A_2({\kappa}'+\kappa))}\over 4{\kappa}'\lambda {\cosh\alpha}, {\cos \beta}}
\nonumber\\
-ik{S_{12}}&=&-ik{{(A_1({\kappa}'+\lambda)-A_2({\kappa}'+\kappa))}\over 4{\kappa}'\lambda {\cosh\alpha} {\cos \beta}}.
\end{eqnarray}
Thus we get
\begin{equation}
{{ik_{-} S_{11}-ik_ 3S_{12}-ik S_{12}}\over {ik_{-} S_{11}-ik_ 3S_{12}+ik S_{12}}}={{\kappa+\lambda+{i\over k}(E\gamma-M{\gamma}')}\over {\kappa+\lambda-{i\over k}(E\gamma-M{\gamma}'})},
\end{equation}
where we have used
\begin{equation}
E\gamma-M{\gamma}'=(E-M)(A_1-A_2).
\end{equation}
Finally, using the relation
\begin{equation}
{{{\kappa-i{{M{\gamma}'}\over k}}\over {\lambda-i{{E\gamma}\over k}}}}={{{{\lambda+i{{E\gamma}\over k}}\over {\kappa+i{{M{\gamma}'}\over k}}}}},
\end{equation}
we get
\begin{equation}
{g^{out}\over g^{in}}={{\hat g}^{out}\over {\hat g}^{in}}
{{\kappa-i{{M{\gamma}'}\over k}}\over {\lambda-i{{E\gamma}\over k}}}.
\end{equation}
\par
One may  calculate the phase shifts by an algebraic technique by observing that the differential equation for $\hat g$ obtained from equation (32) is of the same form as the radial equation for the two-dimensional hydrogen atom with the Hamiltonian
\begin{equation}
H= {{\bf p}^2\over 2}-{\alpha\over r}.
\end{equation}
The latter problem has been treated by Alhassid,Engel and Wu [5].They use the fact that the problem has the dynamical symmetry under the $SO(2,1)$ group so that the asymptotic states obey recursion relations under the action of the generators of the $SO(2,1)$ algebra.On the other hand, the asymptotic states are linear superposition of eigenstates of the Casimir operator of the algebra of the Euclidean group $E(2)$.As a consequence one gets a recursion relation for the ratio of the out and in amplitudes which can be solved. Making appropriate changes in their results:$\alpha \to 2E\gamma $ and $m\to \lambda+{1\over 2}$, we get 
\begin{equation}
{{\hat g}^{out}\over {\hat g}^{in}}=-e^{i\pi\lambda}{\Gamma(1+\lambda-i{E\gamma\over k})\over \Gamma(1+\lambda+i{E\gamma\over k})}.
\end{equation}
Using equation (43) in equation (41) we have the same phase shifts as those of equation (24).
\vskip 1.0 cm
\noindent
{\bf 5. Conclusion}
\par
In conclusion we have shown that the phase shifts for generalized Dirac equation can be found algebraically. The results are in agreement with the conventional calculation.The advantage of the algebraic method is that we work in the $r\to\infty$ limit instead of solving for the radial functions and then taking the limit.
\pagebreak

{\bf References}
\noindent

1. W. Greiner, Relativistic Quantum Mechanics,Springer Verlag,New York,(1990).
\noindent

2 S. I. Ikhadir, O.Mustafa,R. Sever,Hadronic J.{\bf 16}, 57,(1993).
\noindent

3. Q.G.Lin Phys. Lett. A {\bf 260}(1999) 17

\noindent

4. Y. Alhassid,F. G\"ursey and F. Iachello J. Phys.A:Math. Gen.

   {\bf 22}(1989) L947 

\noindent

5.Y. Alhassid,J. Engel and J. Wu Phys. Rev. Lett. {\bf             
       53}(1984) 17 
\noindent

\end{document}